# Graph-Network-Based Predictive Modeling for Highly Cross-Linked Polymer Systems

*Wonseok Lee[1†], Sanggyu Chong[1†], and Jihan Kim[1]*

Department of Chemical and Biomolecular Engineering, Korea Advanced Institute of Science and Technology, 291 Daehak-ro, Yuseong-gu, Daejeon 34141, Republic of Korea

* Corresponding author: J.K. (jihankim@kaist.ac.kr)

ABSTRACT

In this study, a versatile methodology for initiating polymerization from monomers in highly cross-linked materials is investigated. As polymerization progresses, force-field parameters undergo continuous modification due to the formation of new chemical bonds. This dynamic process not only impacts the atoms directly involved in bonding, but also influences the neighboring atomic environment. Monitoring these complex changes in highly cross-linked structures poses a challenge. To address this issue, we introduce a graph-network-based algorithm that offers both rapid and accurate predictions. The algorithm merges polymer construction protocols with LAMMPS, a large-scale molecular dynamics simulation software. The adaptability of this code has been demonstrated by its successful application to various amorphous polymers, including porous polymer networks (PPNs), and epoxy-resins, while the algorithm has been employed for additional tasks, such as implementing pore-piercing deformations and calculating material properties.

## 1. Introduction

Since its inception in the early 1950s, Molecular Dynamics (MD) simulations have undergone continuous development and refinement. Leveraging advancements in computational resources, MD simulations have become widely employed across diverse fields, such as materials science and biochemistry. MD simulations prove particularly advantageous in elucidating phenomena that are challenging or impractical to investigate experimentally due to temporal or circumstantial constraints. However, to effectively conduct MD simulations, it is imperative to define a well-structured model.

In the context of structure determination, crystalline structures offer advantages due to the ease of conducting X-ray diffraction analysis. Employing the Rietveld method[1,2], the initial structure for computational calculations can be determined with considerable accuracy, thereby serving as the foundation for extensive studies of crystalline structures. In contrast, amorphous structures, despite their potential for practical applications, remain largely unexplored due to the challenges associated with their structural determination. Amorphous structures exhibit short-range connectivity while lacking long-range periodic order, resulting in X-ray diffraction patterns characterized by broad, featureless humps.[3] To simulate amorphous materials, a sufficiently large unit cell is required to enable molecular-level observations while minimizing the influence of long-range order. Consequently, quantum computational approaches become impractical for this purpose, and investigations of amorphous materials have predominantly relied on molecular dynamics and Monte Carlo techniques.

Significant advancements in the description of amorphous glassy polymers using Monte Carlo techniques have been made, thanks to the pioneering work of Theodorou and Suter (TS algorithm).[4,5] Monte Carlo-based polymerization algorithms have evolved through various TS-inspired methods, such as configuration-biased algorithms[6-8] and recoil growth algorithms[9-11], ultimately leading to the relatively recent development of random-walk polymerization.[12,13] Several of these techniques can be implemented using commercial programs like Material Studio[14] and MAPS, or open-source codes such as Pysimm[12], stk[15], and Assemble![16]. While there are subtle differences between these algorithms, they primarily function by growing polymer chains to the desired length based on the specific conditions set by each algorithm. However, these algorithms are typically static and they often struggle to generate geometrically complex polymers, as they are limited to forming polymers with simple linear chain-side chain architectures. This highlights the challenges in characterizing highly cross-linked materials, such as PPN and Epoxy-Resin, using Monte Carlo methodologies.

In the realm of polymerization algorithms utilizing Molecular Dynamics (MD), the reactive force-field (ReaxFF)[17], developed for reaction simulation, has become one of the cornerstones of polymerization algorithms, despite not being explicitly designed for self-assembly or the expression of amorphous MOFs. Nevertheless, the ReaxFF methodology is beleaguered by two significant limitations. Firstly, its computational requirements exceed those of conventional Molecular Dynamics (MD) techniques, thereby prolonging the simulation time. Secondly, it is confined to a finite set of atom types for which ReaxFF calculations can be conducted. In light of these constraints intrinsic to ReaxFF, a more traditional yet enduringly favored technique involves an algorithm predicated on distance-based polymerization. Within this sphere of methodology, the Colina group has been at the forefront, with their notable invention, "Polymatic,"[18-22] being the most renowned to date. In addition to the Colina group's contributions, multiple attempts to utilize Molecular Dynamics (MD) simulations have been made in the exploration of PPN and Epoxy-resin.[23-25] Despite the variance in addressed conditions and materials, these algorithms share a fundamental concept: placing monomers in the initial simulation box and connecting them based on distance criteria. However, these algorithms possess a significant limitation: The absence of a facile and rapid force-field assignment system during the connection process. This limitation prevents many MD algorithms from being applied to broader material groups and diverse cases. Even "Polymatic," from a non-expert perspective, appears rather complex in defining the force-field of connection sites. Hence, this study presents a method that can be readily applied to an expanded range of materials by automating intricate force-field calculations and assignments through a graph network.

2. Methods

In conducting molecular dynamics simulations, we employ the Large-scale Atomic/Molecular Massively Parallel Simulator (LAMMPS).[26] LAMMPS is utilized to stabilize the material system following bond formation, as well as to compute various properties upon the completion of system generation.

<A. Algorithm>

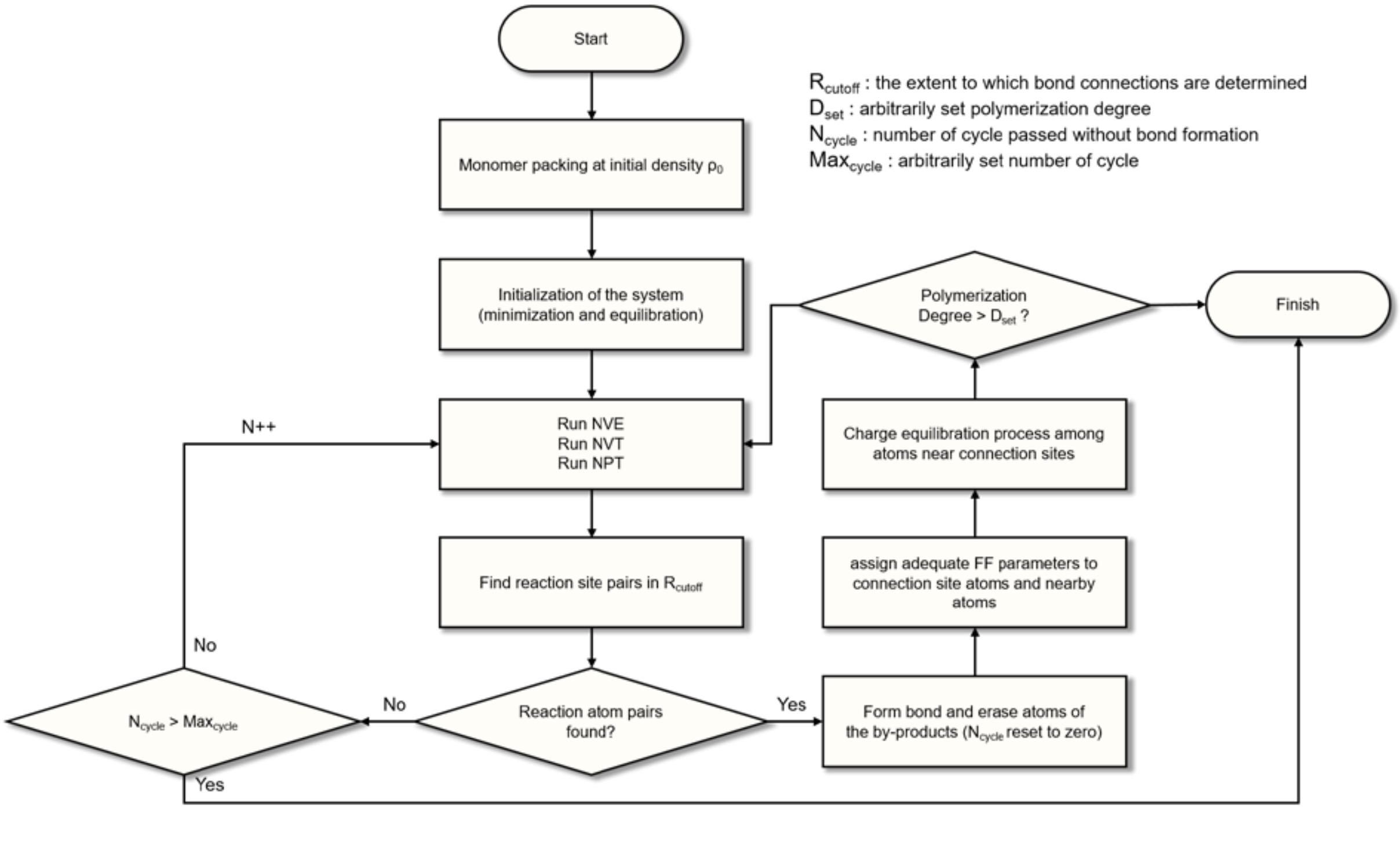

Figure 1. Comprehensive algorithm diagram

In terms of the overall approach, our study shares similarities with "Polymatic" and other polymerization algorithms, with the exception of the force-field parameter application to the structure. The process begins by placing monomers in the simulation box at a very low initial density.[27] This packing of monomers is efficiently executed using the software PACKMOL.[28]

After this initial setup, a recurring loop of bond connection and equilibration ensues. The equilibration steps encompass NVE, NVT, and NPT ensembles, each conducted in carefully tuned cycles tailored to specific material groups. Following this initial setup, a recurring loop of bond connection and equilibration takes place. The equilibration steps encompass NVE, NVT, and NPT ensembles, each conducted in carefully tuned cycles tailored to specific material groups. Frequently, the temperature is set sufficiently high to ensure adequate degrees of freedom in molecular movement.[27] The bond connection process involves the formation of bonds between designated reaction sites in the monomer state. Reaction sites or reaction atom pairs are connected only when they are within the established cutoff (default 5Å) and are the closest reaction sites to each other. The pairs that vanish upon bond formation are tracked and used to calculate the degree of polymerization. The recurring loop is terminated under two primary conditions: one, when the degree of polymerization meets the arbitrarily set target, and two, when the recurring loop has been repeated a certain number of times without any bond formation. These termination conditions can also be adjusted. Once the structure exits the recurring cycle, it undergoes a sufficiently lengthy equilibration process until it is adequately prepared for further simulation.

Regarding force-field parameters, our algorithm currently supports the Dreiding force field[29], Universal force field (UFF)[30], and General Amber Force-Field (GAFF)[31], with plans to incrementally incorporate additional force-field parameters. When applying force-field parameters, we do not focus on intricate details, such as whether specific carbon atoms are in a sp2 or sp3 state. Instead, we vectorize these attributes through a graph network. In scenarios involving low connection complexity and a small number of atoms, this approach does not yield significant differences. However, as the structure becomes more complex, our algorithm's utility increases. Figure 2 provides an overview of the force-field application process within our algorithm.

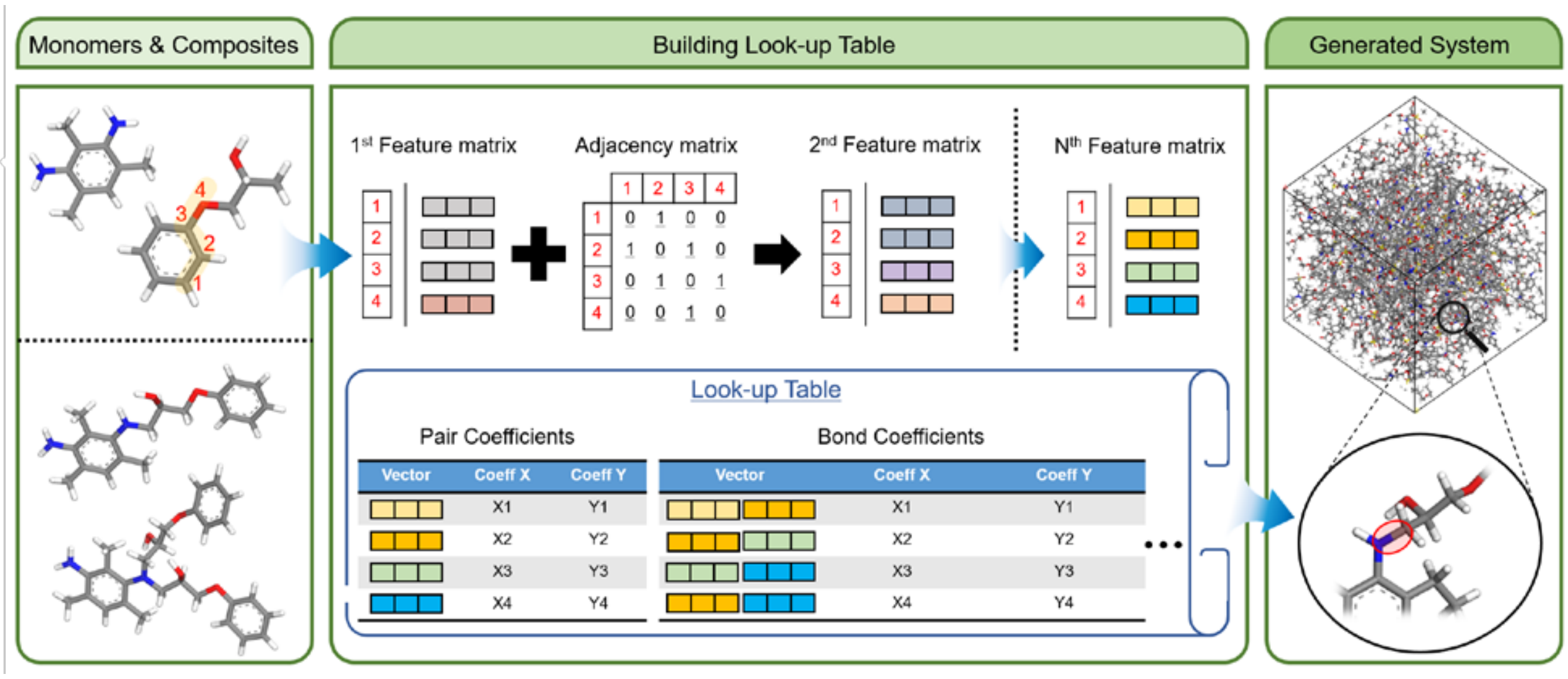

Figure 2. Illustration of creating composites with connection information from monomers and storing their force-fields as vectors.

Upon determining the monomers for polymerization, various complexes that can be formed from these monomers are generated. These may include dimers, trimers, or more complicated structures. It is crucial to ensure that the composites account for all types of connections that may occur during the generation process. This step is performed using the "RDKIT" library[32], with monomers expressed as SMILES representations. To obtain the initial force-field parameters for the monomers and constructed composites, we utilized open-source codes: Lammps Interface[33] for acquiring UFF and Dreiding force fields, and AmberTools[34] for the GAFF force field. Atoms of monomers and composites are vectorized based on the element type and each atom's adjacency matrix. The vectorized forms of atoms are then matched with the obtained force-field parameters to create a "Look-up" table. In other words, force-field parameters are derived from an existing simple system, stored, and subsequently applied to a more complex system.

<B. Details>

-

When dealing with bent monomers, the system can occasionally cluster together, consuming all the reaction sites and forming dimers. To prevent this, we imposed limitations on bond formation between reaction sites belonging to adjacent frameworks, as well as restricting the number of bonds that can form in a single cycle. Furthermore, we designed the system to automatically remove byproducts, such as hydrogen and halogenated gases, which are frequently generated during reaction periods.

When creating the aforementioned "Look-up" table of force-field parameters, each atom of the monomers and composites is treated as a node in the graph. Initially, each atom is assigned a vector based on its element type. Subsequently, the vector is updated through the vectors of other closely connected atoms, according to the expression detailed below.

As the updates progress, the vector features of each node (or atom) become more distinct, and with continued updating, every atom can be differentiated from one another, barring those at a perfect symmetry point. In our approach, we only perform updates to the level of uniqueness required by a specific force-field, typically necessitating 4 to 6 iterations.

## 3. Results and discussion

The different molecular systems discussed below are generated using the aforementioned code. Appropriate charge and force-field parameters are selected to match the characteristics of each material. Detailed conditions, such as the number of cycles and system size, are also finely tuned.

<A. Epoxy-Resin cross-linked polymer>

Epoxy-resin materials are synthesized through the reaction of epoxies, which possess varying numbers of epoxide groups, and hardeners that facilitate bonds between epoxies, resulting in highly cross-linked substances. In this study, hardeners with amine groups are employed, and the ring structures of epoxide groups are prepared in their activated form, as described by Li et al.[35] All monomer units undergo a minimization process using the Material Studio software before being introduced into the simulation box. Monomer units utilized in this study to create cross-linked epoxy-resin polymers are presented in Table 2, while the potential reactions between them are summarized in Figure 3.

| Material | Chemical structure |
|---|---|
| DGEBA | 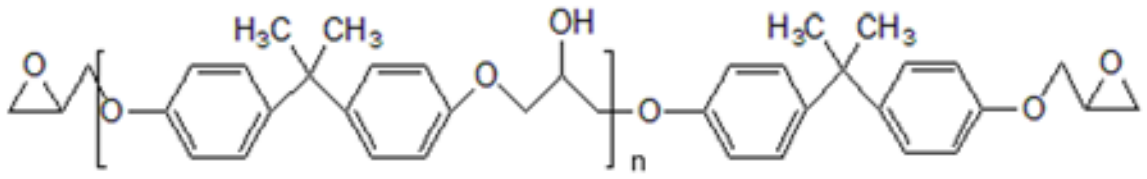 |
| TGDDM | 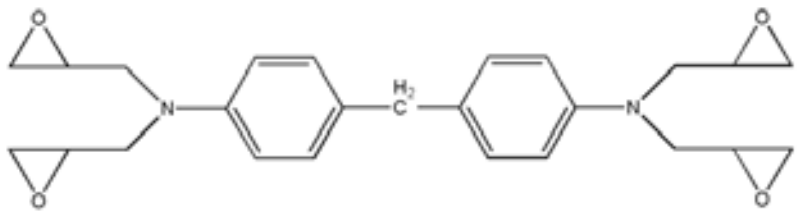 |
| TMBP | 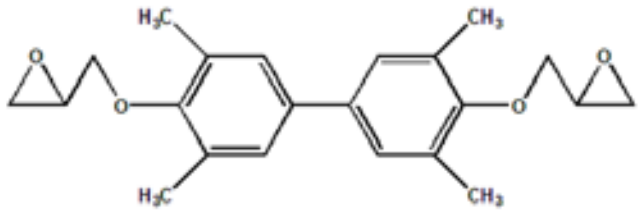 |
| TGPAP | 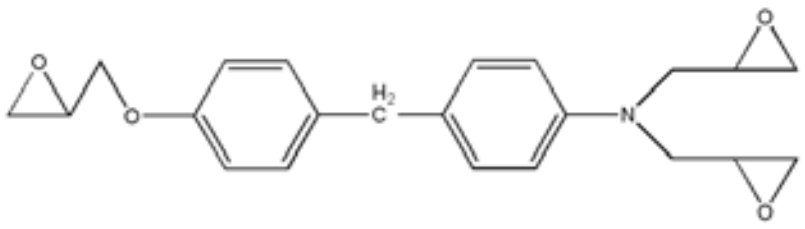 |
| DGEBF | 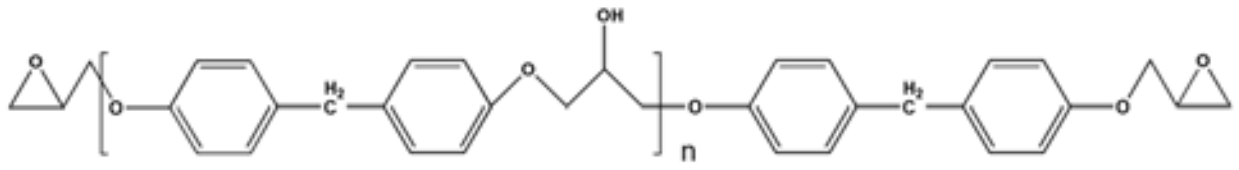 |
| IPD | 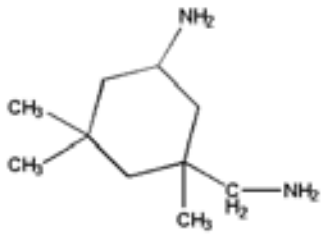 |
| TETA | 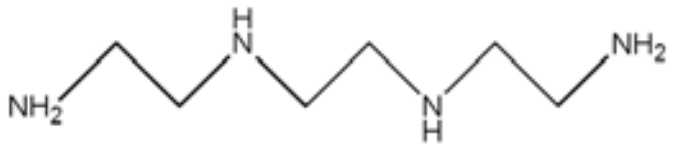 |
| PDA | 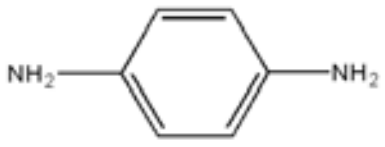 |
| DDS | 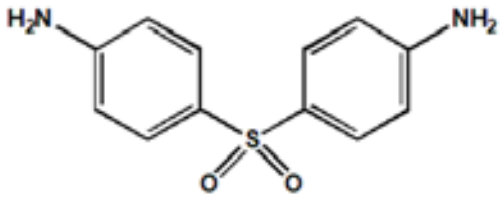 |

DGEBA: Diglycidyl ether bisphenol A
TGDDM: Tetraglycidyl-4,4'-diaminophenylmethane
TMBP: Tetramethylbiphenyl diglycidyl ether
TGPAP: Triglycidyl-p-aminophenol
DGEBF: Diglycidyl ether bisphenol F
IPD: Isophorone diamine
TETA: Triethylenetetramine
PDA: p-phenylenediamine
DDS: 4,4'-Diaminodiphenyl sulfone

Table 2. A list of the monomer units that were utilized in this study to computationally synthesize cross-linked epoxy resin polymers

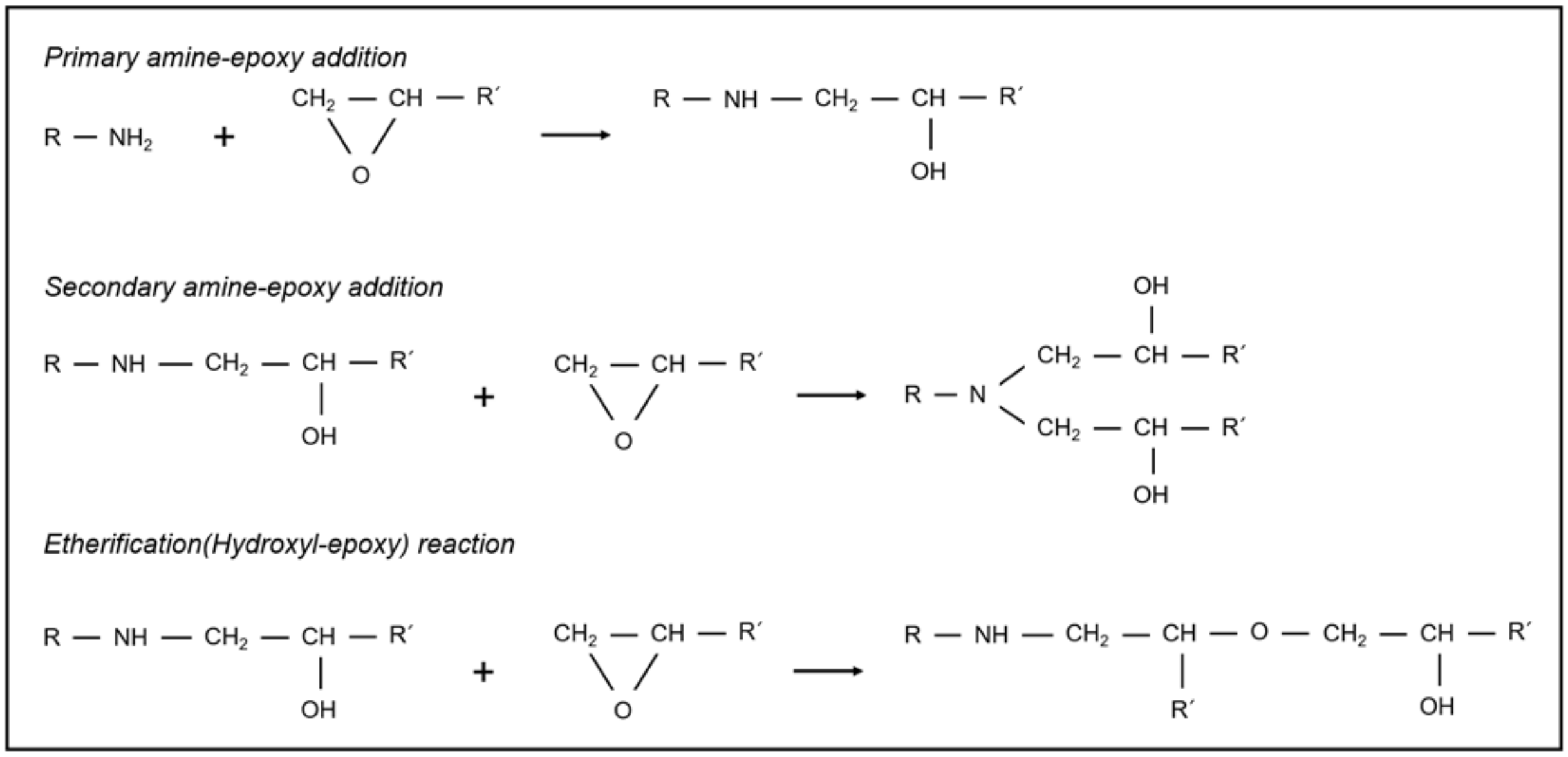

Figure 3. A compilation of the reactions that occur during the polymerization of epoxy-resin materials.

In this study, we consider not only the primary and secondary amine reactions, which are the main reactions in the production process of epoxy-resin, but also etherification reactions. This comprehensive approach is critical as maximizing the degree of conversion is essential for optimizing certain properties, such as Tg, a finding underscored by Barton's research[36]. The amine reaction is implemented with priority when it conflicts with the etherification reaction, meaning that the etherification reaction occurs only after all primary and secondary amine reactions have terminated. After testing various force-field parameters and atomic charges, we have settled on parameters from the General Amber Force-Field (GAFF) and partial charges from the semi-empirical method with bond charge correction, AM1-BCC[37,38]. Additionally, dielectric constants are set to 78, which corresponds to the value of water, specifically for epoxy-resin materials to ensure better representation.

To validate our computationally constructed epoxy-resin structures, glass transition temperature (Tg) and their mechanical property values are calculated and compared with experimentally measured values.[39-42] For Tg measurement, density is measured every 10K while annealing the system from high to low temperatures. Subsequently, Tg is determined through the density-temperature relationship using piecewise regression. In the case of mechanical properties, it was derived through the Voigt matrix.

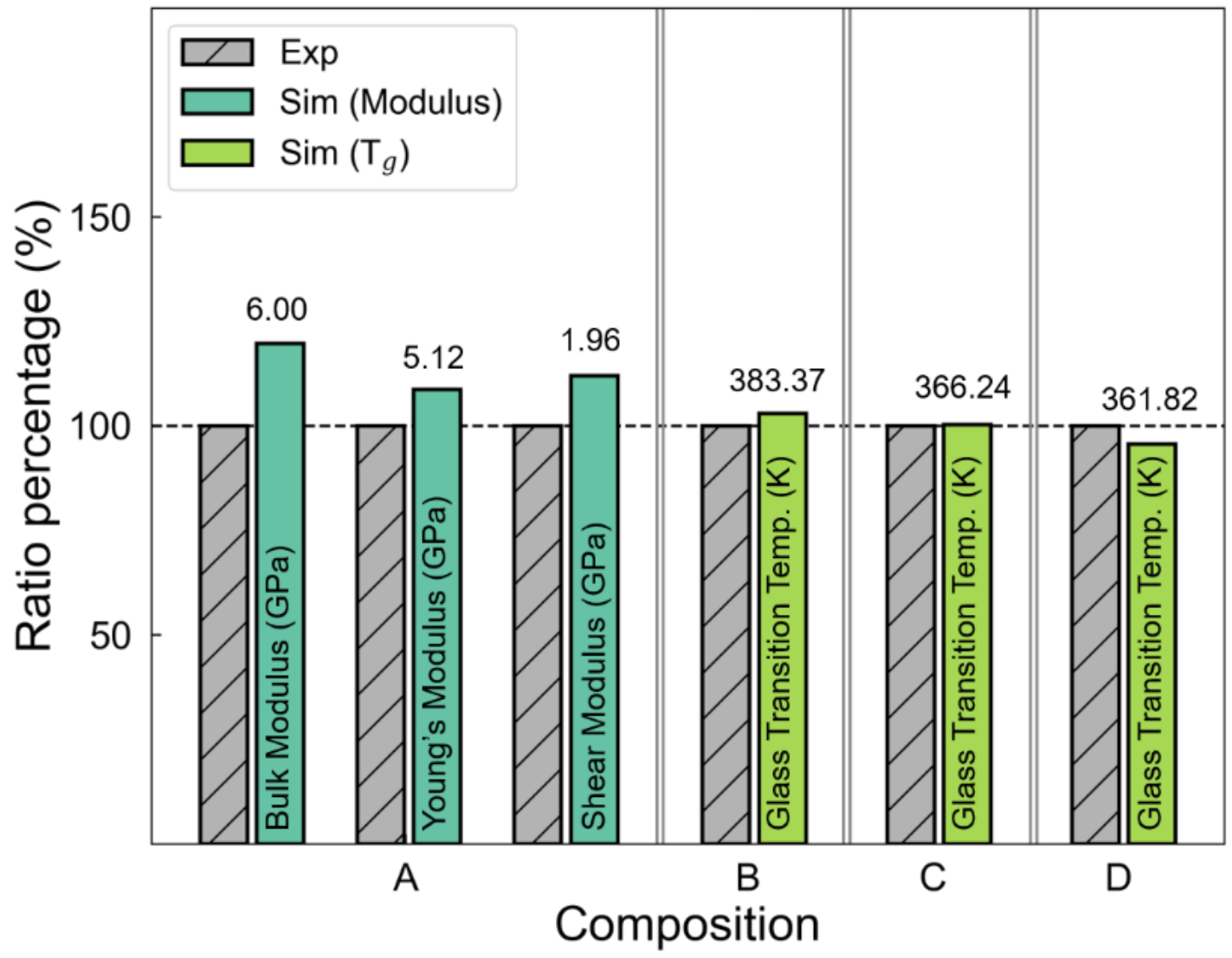

Figure 4. Calculated density and glass transition temperature for compositions A through D. The y-axis shows the calculated value as a percentage of the experimental value.

| Monomer Units | Composition (wt%) | | | |
|---|---|---|---|---|
| | A | B | C | D |
| DGEBA | 80.179 | - | - | - |
| TMBP | - | 76.824 | 92.987 | - |
| DGEBF | - | - | - | 86.655 |
| IPD | 19.821 | - | - | - |
| PDA | - | 23.176 | 7.013 | - |
| TETA | - | - | - | 13.345 |
| Reference | [39] | [40] | [40] | [41] |
| Total number of atoms (#) | 24850 | 24940 | 24984 | 24928 |

Table 3. The composition ratio expressed as the weight percentage (wt%) of each epoxy and hardener.

Next, epoxy-resins of four different compositions are generated computationally and their characteristics are compared. Since commercially purchased monomers are used as weight percentages in experiments, they are converted to molecular counts to simulate the same environment. When the monomer has a repeating unit, we adjusted the number of repeating units from n = 0 to n = 2 to match the epoxy equivalent of commercial epoxy. The composition ratio expressed in monomer units can be found in Table 3. All systems are designed to contain approximately 25,000 atoms. While Tg results are well-aligned with the experimental values, calculated modulus results tend to be higher than experimental values. However, this variation is quite notable, considering that mechanical modulus values in experiments can vary based on the experimental conditions. Therefore, achieving overall inter-material tendencies in good agreement with these fluctuating experimental values, even with slight differences, is a noteworthy accomplishment in our simulation methodology.

Moreover, the correlation between physical properties and the extent of hardening was scrutinized. In prior investigations by Xie et al[43] and Jeyranpour et al[44], the density, Tg, and Young's modulus of epoxy polymers exhibited an increase concomitant with the rise in cross-link density, while the coefficient of thermal expansion (CTE) demonstrated a decrease. To examine the applicability of these tendencies using the methodology employed in this study, the aforementioned four properties were computed by solely modulating the cross-link density while maintaining a constant composition ratio. An epoxy resin system composed of bisphenol A diglycidyl ether (DGEBA; n=0) and 4,4'-diaminodiphenyl sulfone (DDS) was selected for analysis. Although the standard deviations for Tg and Young's modulus were marginally elevated, Figure 5 demonstrates that the trends align with those observed in prior research. The outcomes were obtained by generating five individual systems via distinct random seeds and subsequently averaging the results.

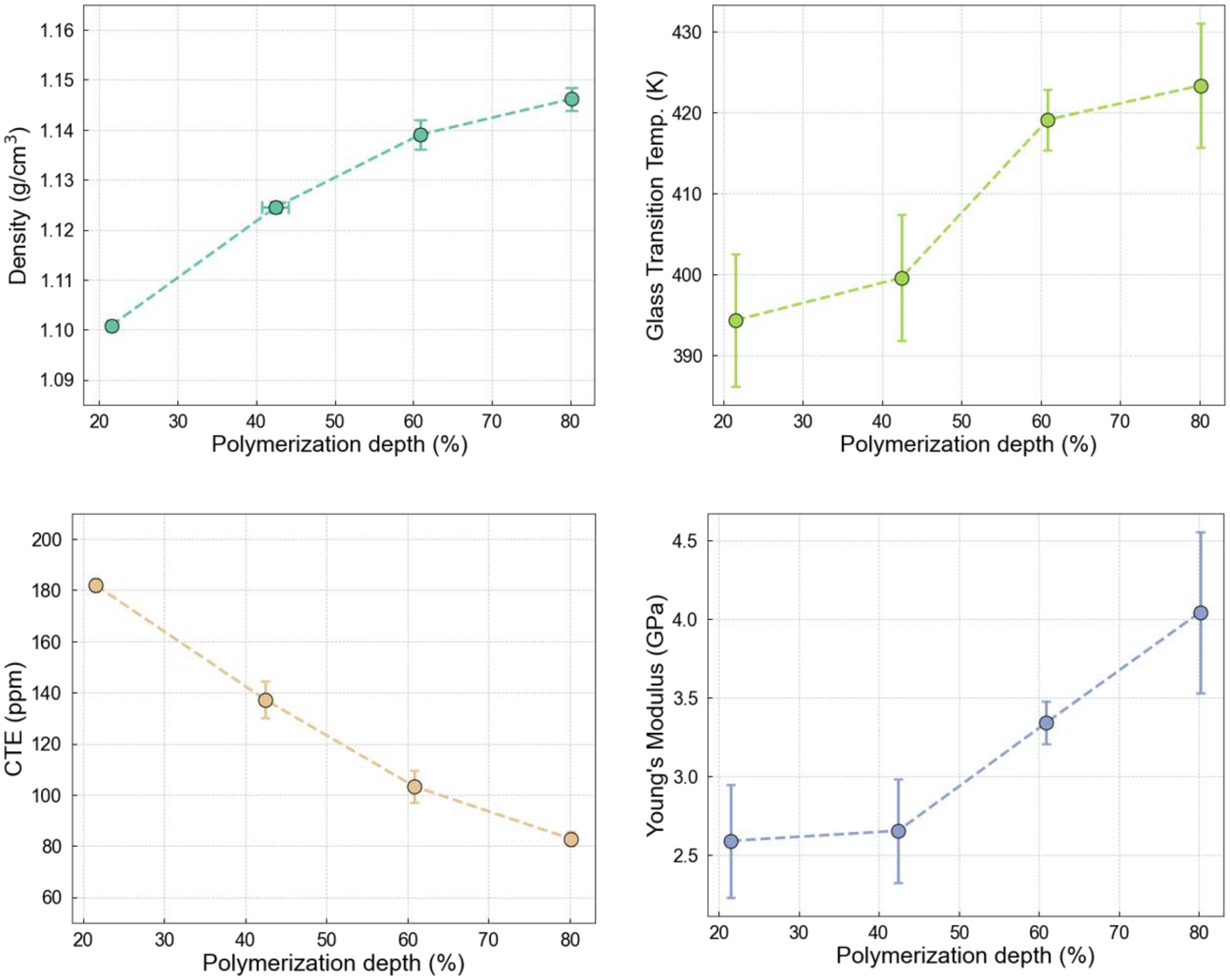

Figure 5. Calculated mechanical properties of an epoxy resin system composed of DGEBA and DDS at varying polymerization depths

An additional benefit of the proposed methodology is its capability to generate a structure comprising multiple monomer unit types simultaneously. This is particularly noteworthy as, while there are many simulation studies focused on single-kind epoxy with single-kind curing resin, there are only very limited studies about systems using multiple kinds of epoxy in a single setup. Our approach addresses this gap in research, exploring the complexities of multi-epoxy systems. For example, an epoxy polymer containing three different epoxy monomer units was synthesized and illustrated in Table 4 and Figure 6. Analogous to preceding procedures, the structure was established with computational means and compared with experimental[42].

| Monomer Units | Kind | Weight percent (%) | Molecular quantity |
|---|---|---|---|
| TGPAP | Epoxy | 27.54 | 170 |
| DGEBF | Epoxy | 7.87 | 43 |
| TGDDM | Epoxy | 27.54 | 111 |
| DDS | Hardener | 37.06 | 261 |
| Total | - | 100.01 | 585 |

Table 4. Weight percentages and molecular quantities for each monomer included in the complex composition tested in the study. The chemical composition was referred to in reference [42]. Catalyst not specified

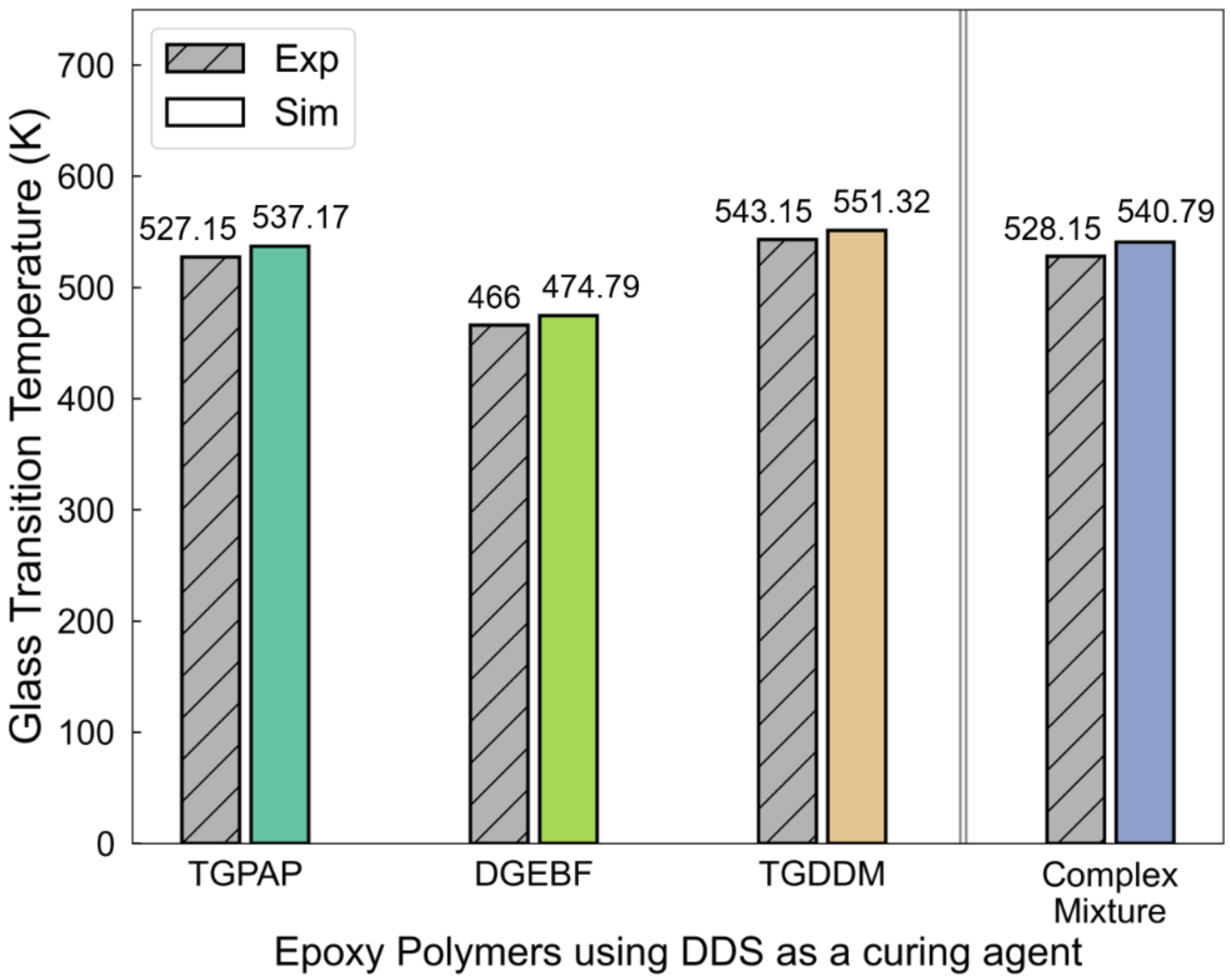

Figure 6. Calculated glass transition temperature of the complex composition studied

Table 4 presents a summary of the weight percentages and corresponding molecular quantities for each monomer. The integrity of epoxide groups in epoxy monomers may be compromised depending on manufacturing and storage conditions. This factor was taken into account when converting weight percentages to molecular quantities, with missing epoxide groups being substituted by hydrogen.

In the results, it can be seen that not only the Tg of single-monomer kind epoxy polymers but also that of multi-monomer kind epoxy polymers are well matched. This not only demonstrates the accuracy of the algorithm we have developed but also proves its easy applicability to multiple monomer kinds, and further, to various materials. In the example above, we only used a total of 4 distinct monomers, including the curing resin, but it is also possible to mix a larger number of monomers. In such cases, it is expected that a larger unit cell size and a greater number of atoms will be required to ensure adequate use of each epoxy.

< B. Porous Polymer Networks>

In the computational synthesis of porous polymer networks (PPNs), we closely adhered to the methodology employed for creating epoxy-resin materials. However, to enhance the robustness of the connection points in PPNs, the dihedral coefficient was derived from the DREIDING force field. Given that numerous studies have focused on measuring pore volume and surface area in PPN materials[45-50], we computed both properties for the PPN-6 material and presented results for polymerization degrees surpassing 50%.

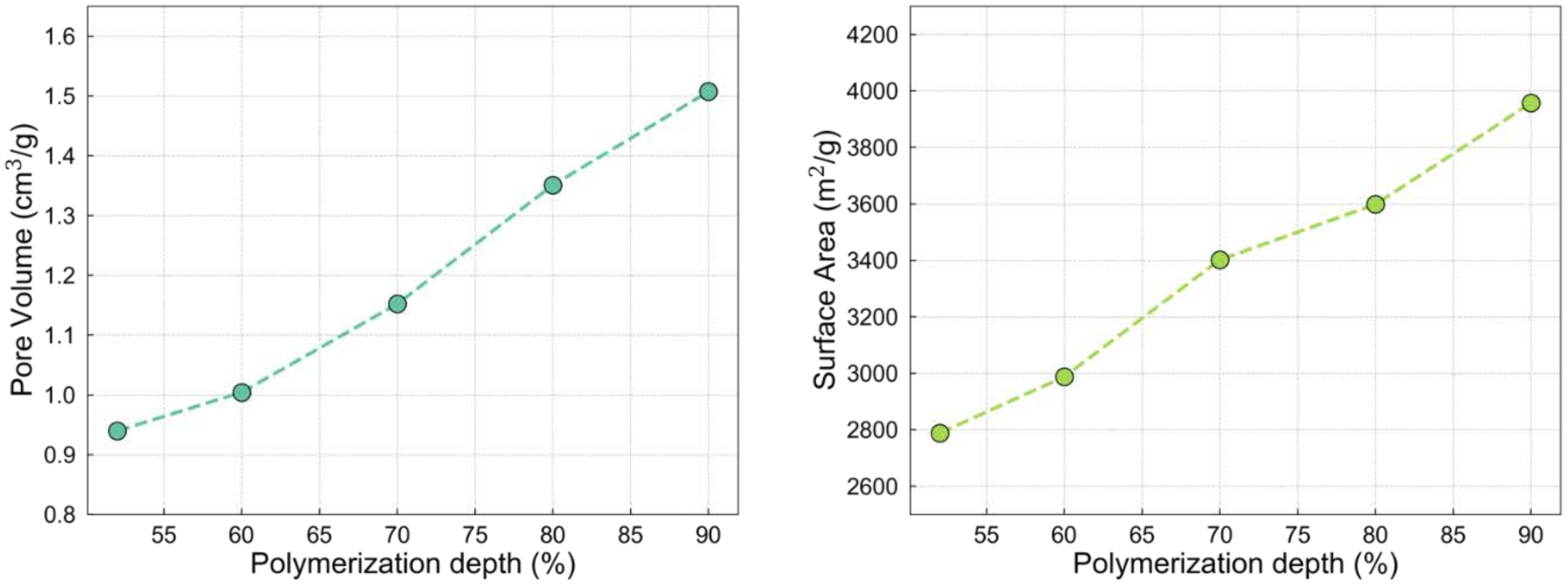

Figure 7. The relationship between the degree of polymerization and the pore volume and surface area of PPN-6

As depicted in the Fig 7, the pore volume and surface area generally exhibit a positive correlation with the degree of polymerization. Typically, but not invariably, PPNs and their analogs exhibit this trend due to the formation of a greater number of pores with diverse sizes as the degree of polymerization increases—a distinctive attribute of the PPN family of materials. These observations indicate that our computational algorithm effectively discerns the variations among different material families.

### < C. Post-processing routine to create polymeric phases>

Frequently, polymers of interest for modeling undergo phase transitions and form unique structures at various length scales, rendering the as-generated bulk phase an inadequate representation. This is particularly relevant when the polymer is involved in the templated synthesis of hierarchically porous media, where host-guest interactions between the polymer matrix and other guest molecules are of interest. To accurately model these systems, it is essential to construct slabs or introduce spherical and cylindrical pores within the as-generated bulk phase polymer. Consequently, a post-processing code has been further developed to facilitate the construction of such slab or porous systems, thereby providing a more accurate representation of the real polymeric systems under investigation.

To create such structures, sections of the unit cell must be excised from the generated bulk polymer phase to generate voids with the desired geometries. In the case of a slab, all portions of the polymer within a specified range along one of the unit cell axes can be removed. For the creation of spherical or cylindrical pores, parts of the polymer within a cylinder or sphere of the desired radius can be removed. When performing the partial removal of the polymeric phase, it is crucial to ensure that the remaining polymer structure maintains chemical stability. To address this, our post-processing routine was designed to execute the removal process on a monomer-by-monomer basis, thus preventing unrealistic monomer fragmentation. Another consideration is the exposure of dangling bonds following partial removal from the bulk polymeric phase. Our post-processing routine accounts for this by tracking all dangling bond sites and subsequently "capping" the desired terminal groups onto these sites in each scenario. Following the capping process, energy minimization is performed to guarantee that the templated polymeric structure and the appended terminal groups at the dangling bond sites are described in an energetically stable manner.

## 4. Conclusion

The potential for future research, particularly in the underexplored area of amorphous materials, is exciting. The algorithm's accessibility and flexibility to work with various amorphous polymers make it a valuable tool not only for expanding the material space for amorphous materials but also for researchers seeking to understand the complex behavior of highly cross-linked materials during polymerization.

The efficacy of this approach was demonstrated through the application to PPN and Epoxy-resin polymers. To gain a deeper understanding of the general behavior of amorphous materials, it would be valuable to apply this approach to a broader range of materials. To achieve greater stability in the results, the unit cell size could be increased to accommodate a larger number of atoms, or multiple structures could be generated and their averages calculated.

## AUTHOR INFORMATION

**Corresponding Author**

Jihan Kim − Department of Chemical and Biomolecular Engineering, Korea Advanced Institute of Science and Technology (KAIST), Daejeon 34141, Republic of Korea;

Email: jihankim@kaist.ac.kr

**Author Contributions**

Wonseok Lee − Department of Chemical and Biomolecular Engineering, Korea Advanced Institute of Science and Technology (KAIST), Daejeon 34141, Republic of Korea

orcid.org/0000-0002-6007-0495

Sanggyu Chong − Department of Chemical and Biomolecular Engineering, Korea Advanced Institute of Science and Technology (KAIST), Daejeon 34141, Republic of Korea

Present address: Laboratory of Computational Science and Modeling, Institute of Materials, École Polytechnique Fédérale de Lausanne, 1015 Lausanne, Switzerland

orcid.org/0000-0002-6948-1602

ACKNOWLEDGMENT

This work was supported by KOLON INDUSTRIES, INC.

All authors have contributed equally.